\documentclass[10pt,letterpaper]{article}

\usepackage{cogsci}
\usepackage{pslatex}
\usepackage{apacite}
\usepackage{graphicx}
\usepackage{subcaption}
\usepackage{makecell}
\usepackage{multirow}
\usepackage{multicol}

\cogscifinaltrue

\title{How do people watch AI-generated videos of physical scenes?}
 
\author{{\large \bf Danqing Shi$^{1,\dagger}$, 
Lan Jiang$^{1}$,
Katherine M. Collins$^{1,2}$, 
Shangzhe Wu$^{1}$, 
Ayush Tewari$^{1}$, 
Miri Zilka$^{1,\dagger\dagger}$
} \\ 
  University of Cambridge$^{1}$, Massachusetts Institute of Technology$^{2}$ \\
  \{$^{\dagger}$ds2206, $^{\dagger\dagger}$mz477\}@cam.ac.uk
}

\begin{document}

\maketitle

\begin{abstract}
The growing prevalence of realistic AI-generated videos on media platforms increasingly blurs the line between fact and fiction, eroding public trust. 
Understanding how people watch AI-generated videos offers a human-centered perspective for improving AI detection and guiding advancements in video generation.
However, existing studies have not investigated human gaze behavior in response to AI-generated videos of physical scenes.
Here, we collect and analyze the eye movements from 40 participants during video understanding and AI detection tasks involving a mix of real-world and AI-generated videos. 
We find that given the high realism of AI-generated videos, gaze behavior is driven less by the video’s \textit{actual} authenticity and more by the viewer's \textit{perception} of its authenticity. Our results demonstrate that the mere awareness of potential AI generation may alter media consumption from passive viewing into an active search for anomalies.

\textbf{Keywords:} 
Generative AI;
Eye Tracking; 
Gaze Analysis;  
Video Generation;
Human-AI Interaction 
\end{abstract}
\section{Introduction}

``\textit{If you can't tell the differences, does it matter?}'' was a line from \textit{Westworld} (TV Series, 2016–2022) in response to the question (to an AI) ``Are you real?''. This question is becoming increasingly relevant to our reality. 
The widespread adoption of AI-generated videos has been facilitated by the rapid development of video generation models such as Sora~\cite{sora} and Veo~\cite{google2025veo3}, enabling video creation through simple text prompts or image inputs.
People engage with AI-generated video content, but simultaneously express concerns about being deceived, as evidenced by frequent fact-checking behaviors on social media, such as asking ``@grok is this true?'' on X (formerly Twitter)~\cite{renaultgrok} and whether or not a video (or anything) is AI-generated~\cite{isthisAI}.

Yet, it is not clear what the impact of AI-generated content is on human behavior --- and whether people can reliably detect it.  To improve our understanding, we look beyond the final decision of ``real or fake'' and examine the moment-by-moment visual processing that drives these judgments. 
Eye-tracking technology provides a powerful data-driven approach for studying human behavior and cognitive processes, by measuring where a person is looking, for how long and in what sequence, as established in previous cognitive science research~\cite{teng2025eye, teng2024eye, hsiao2020role, tanaka2019utilizing}. 
By measuring visual attention, it gives us a window into human cognitive processes that complements conscious self-reporting (e.g., surveys).  Moreover, this gaze behavioral data has the potential to help define human-centered metrics for evaluating AI-generated videos, and drive more targeted improvements for better video generation models in practice.

\begin{figure}[!t]
\centering
\includegraphics[width=1.0\linewidth]{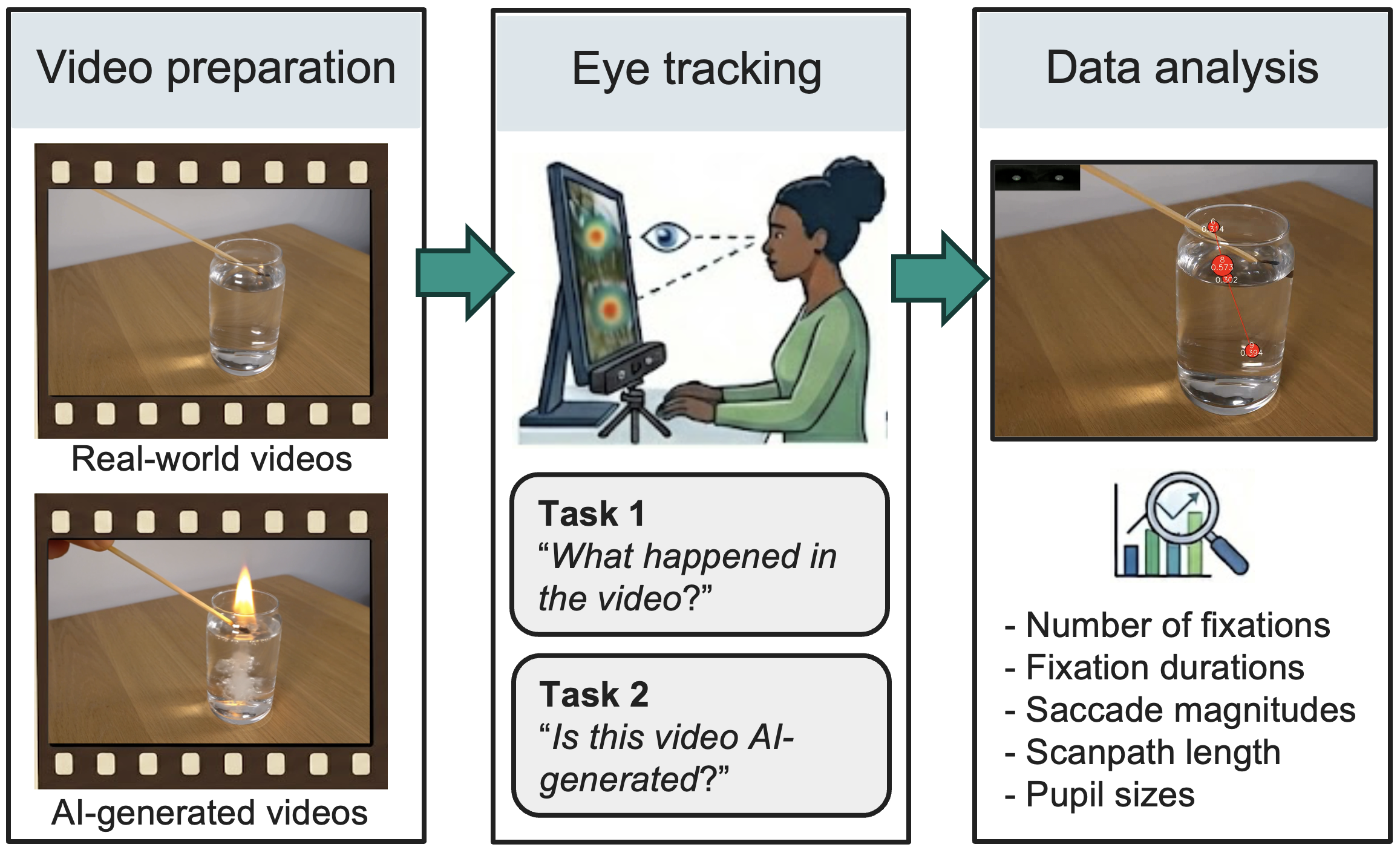}
\caption{This study investigates human gaze behavior from 40 participants with diverse backgrounds when watching real-world and AI-generated videos of physical scenes. Participants are asked to watch videos normally for understanding or to detect whether the videos are AI-generated or not. Eye tracking data is collected during the experiments for analysis.} 
\label{teaser}
\vspace{-3mm}
\end{figure}

Previous eye-tracking studies have investigated differences in cognitive processing using gaze data when comparing the reading of human-written and AI-generated text~\cite{ilyas2025reading},
proofreading behavior with and without AI assistant~\cite{shi2024crtypist},
human visual perception when individuals view partially forged images compared to authentic images~\cite{cartella2024unveiling},
and human perception of real and fake face images~\cite{huang2025analysis}.
Generally, these results in static content show that when people view AI content, their attention tends to focus on smaller and more targeted regions, as the AI content might be more predictable in terms of attention. 
Concerning videos, existing studies have mainly explored face-swapped videos~\cite{gupta2020eyes, wohler2021towards}.
The analysis of eye-tracking patterns showed differences in fixation behavior, with participants focusing more on the mouth area, which is usually unnatural during speech in face-swapped videos. 
However, the most common generation technique at that time was DeepFake~\cite{westerlund2019emergence}, which is much less advanced than recent generative AI models. 
Furthermore, the existing work only studied human videos. The relationship between human behavior and cognitive processes during engagement with real versus AI-generated videos of physical scenes remains insufficiently understood. 
As world models improve in understanding and generating complex physical phenomena~\cite{motamed2025generative}, understanding how people evaluate the authenticity of physical scenes becomes especially important. 

This paper aims to investigate how people watch AI-generated videos of physical scenes, without featuring real or AI-generated human faces.
Two main tasks are considered in the experiment: (1) video understanding, where individuals normally watch videos and focus on the content, regardless of whether it is real or AI-generated, and (2) AI detection, where viewers are asked to identify which video is real and which is not. In both tasks, participants view a mix of real and AI-generated videos. A total of \textit{21,379 fixations} across \textit{1,573 scanpaths} and \textit{800 human responses} were recorded from 40 participants watching 80 videos (40 AI and 40 real videos).
The study dataset is released to support future research~\footnote{Study data can be found at https://github.com/sdq/gaze-genai}.

Intriguingly, we find that gaze patterns did \emph{not} vary between real and AI-generated videos. However, we did find differences in viewing patterns between the two tasks: Participants' eye movements were more actively gathering information during AI detection, but spent less effort on each point. Moreover, although the authenticity of a video did not affect viewing behavior, human gaze patterns did vary based on the participants' \emph{perception} of whether or not the video was real. Participants sample more information and put more effort into the videos they ultimately consider real. 
We also found differences in gaze behaviors based on a qualitative analysis of participants' self-reported strategies for AI-video detection, which are collected through user questionnaires after the experiment. 
The result shows that people who reported a logical strategy for detection show a more consistent eye movement behavior than those relying on intuition. 
Our work highlights the value of leveraging techniques from cognitive science toward naturalistic problems of pressing importance: whether people can reliably detect the increasing plethora of AI-generated content circulating on the web --- and what the impact of that increased content is on how people consume such content. 



\section{Method}

We next overview our experimental design. 





\subsection{Experimental Design}

We conduct an eye-tracking study to investigate everyday video-watching behavior when viewing real-world videos and AI-generated videos, testing the hypotheses.
The eye movement data of all participants is recorded during video watching.
All methods were carried out in accordance with the relevant guidelines and regulations of the University of Cambridge.
All experimental protocols have been reviewed and approved by the Department of Engineering Ethics Committee, University of Cambridge.

\subsubsection{Participants}

We recruited $N{=}40$ adults (26 females, 18–48 years old, 27.1 in average), with normal or corrected-to-normal vision. 
Participants who wear glasses or contacts were allowed.  
Two of the participants reported that they have slight ADHD (P3 and P39).
40 participants are recruited from interdisciplinary fields, including computer science (32.5\%), social sciences and humanities (27.5\%), engineering (17.5\%), natural sciences (7.5\%), and other professional areas such as MBA and law (10\%).
More than half of the participants (60\%) had no experience of using AI videogen tools.
Each participant received 15 GBP in compensation for the experiment, which lasted approximately 30 minutes.
Informed consent was obtained from all subjects.

\begin{figure}[!t]
\centering
\includegraphics[width=1.0\linewidth]{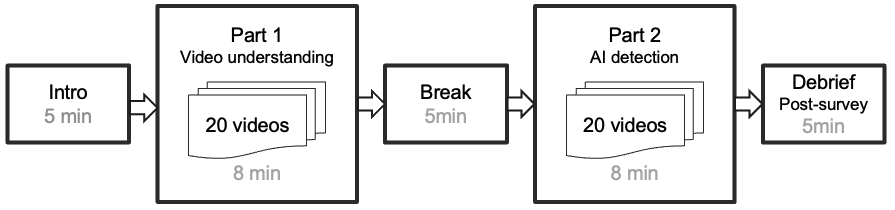}
\caption{Study procedure} 
\label{procedure}
\vspace{-3mm}
\end{figure}

\subsubsection{Materials}

Two sets of video stimuli were prepared: 40 real-world and 40 AI-generated videos, resulting in a total of 80. 

\textit{Real-world videos.}
Real-world videos include two data sources: Physics-IQ dataset (S1)~\cite{motamed2025generative} and Adobe Stock (S2)~\cite{adobestock}.
Each physics from the Physics-IQ dataset~\cite{motamed2025generative} video depicts a distinct physical scenario. All videos are recorded at 30 frames per second.
We picked 20 videos depicting different scenarios from five categories: Solid mechanics, Fluid dynamics, Optics, Thermodynamics, and Magnetism.
To increase the diversity of the video set, we also selected professionally edited videos from Adobe Stock~\cite{adobestock}, which offers high-quality labels that distinguish between human-made and AI-generated content.
We picked 20 human-made videos from four general categories to cover diversity: nature, wildlife, food and drink, and sports. We ensure there are no identifiable humans in the videos.


\textit{AI-generated videos.}
Comparable AI-generated videos were generated from these real-world videos. 
To ensure the similarity between real and AI-generated videos, the text description in the prompt was used to ensure reproduction of the same physical phenomenon (e.g., ``\textit{A lit match is being lowered into a glass of water.}'' for the video illustrated in Figure~\ref{teaser}), while the keyframe was included as a part of the prompt to maintain the same video style.
For the physics videos, the keyframe for each video was selected to provide enough information about the physical event and objects~\cite{motamed2025generative}.
For the professional videos, the first frame was selected for each video.
Google Veo 3.1 (October 2025)~\cite{google2025veo3} served as the generative model for creating AI videos. 
All real-world and AI-generated videos were cropped to 960 x 720 pixels without any AI-generation marks. Videos were manually trimmed to five seconds to match the length and motion of the real videos.

\subsubsection{Apparatus}

We conducted the experiment with a Gazepoint GP3 eye-tracker, 60Hz system and no head mount. 
The videos were shown on a monitor with 1680 x 1050 px resolution, at 90 ppi (Samsung SyncMaster 226aw). 
Participants were instructed to sit about 60-70 cm from the monitor. Before the experiment started, participants could adjust the seat and the monitor to match their comfortable settings.
We used Gazepoint Control software for calibrating the eye-tracker and Gazepoint Analysis software to show the stimulus videos and record data from participants.


\subsubsection{Procedure}

The study procedure has two parts (see Figure~\ref{procedure}). We decompose the study to assess to kinds of watching behavior: understanding the video and then detecting whether the video was AI-generated. AI-generated and real videos were randomly counter-balanced across the two parts of the experiment. For each scene (e.g., leopard on the snow), each participant only sees one video, either AI or real. Therefore, participants do not make a direct comparison between real and AI-generated videos depicting the same scene. 

\begin{figure*}[!t]
\centering
\begin{subfigure}[t]{0.9\textwidth}
    \centering
    \includegraphics[width=0.9\textwidth]{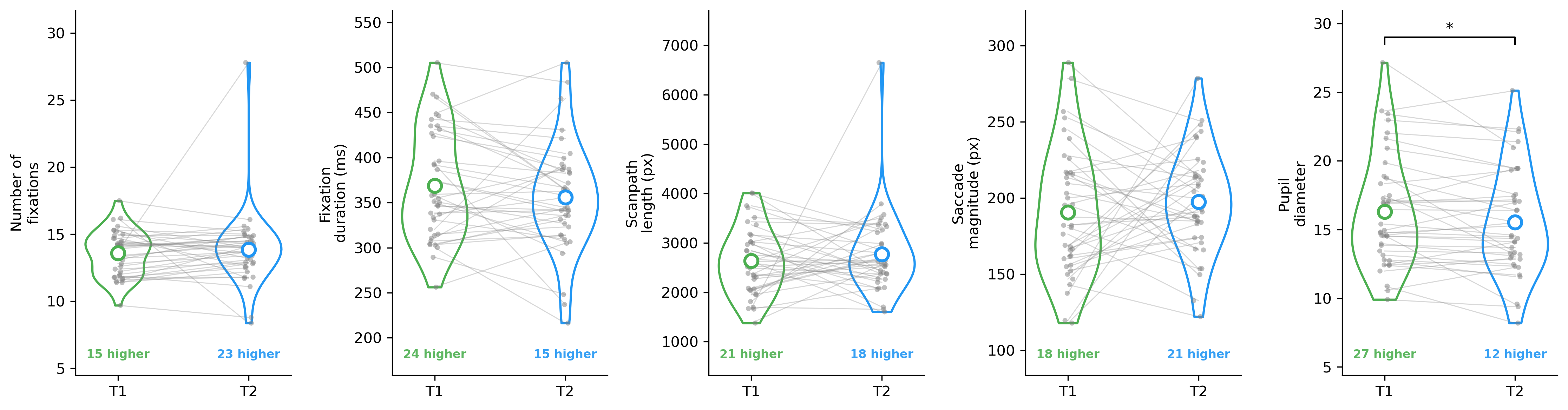}
    \caption{When watching real videos, participants show larger pupil size (*) during normal watching (T1) than AI detection (T2).}
    \label{analysis-a}
\end{subfigure}
\begin{subfigure}[t]{0.9\textwidth}
    \centering
    \includegraphics[width=0.9\textwidth]{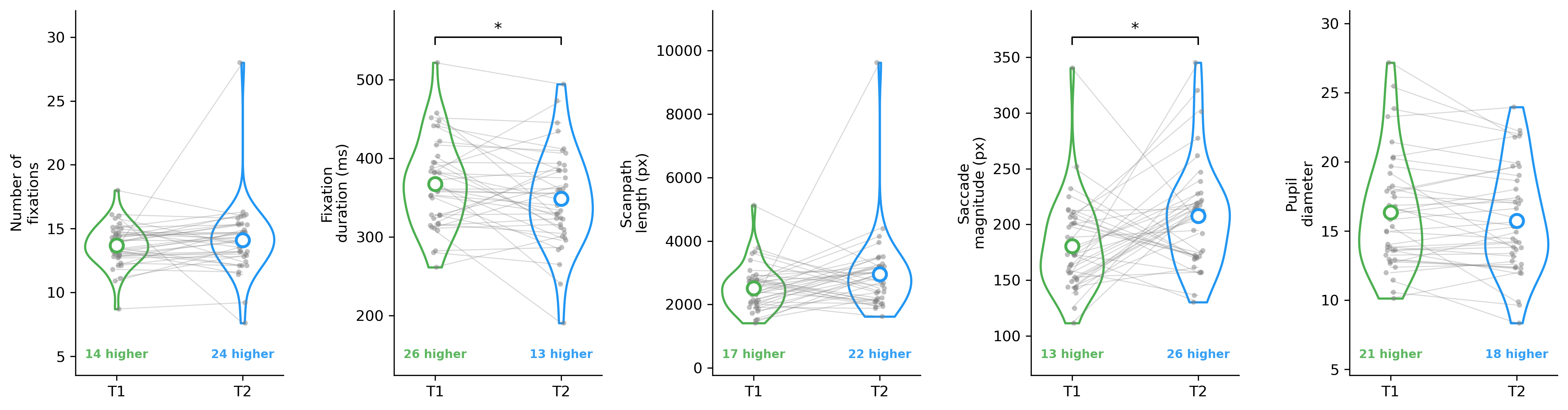}
    \caption{When watching AI-generated videos, participants had shorter fixation durations in each place (*) but longer saccade magnitudes (*) during AI detection (T2).}
    \label{analysis-a}
\end{subfigure}
\begin{subfigure}[t]{0.9\textwidth}
    \centering
    \includegraphics[width=0.9\textwidth]{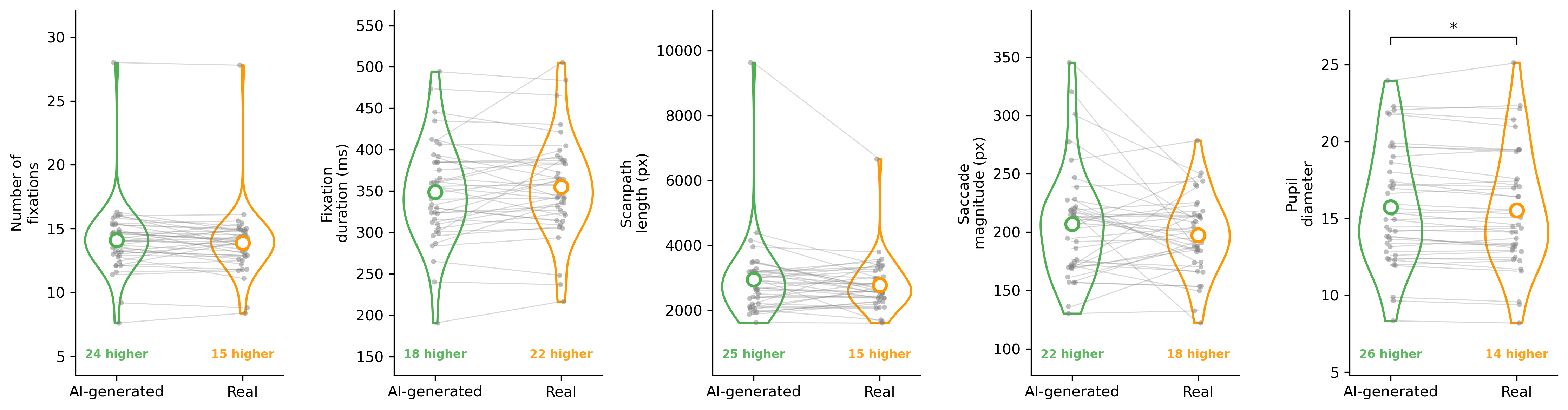}
    \caption{In AI detection tasks, participants' pupils dilate (*) when evaluating AI-generated videos.}
    \label{analysis-b}
\end{subfigure}
\begin{subfigure}[t]{0.9\textwidth}
    \centering
    \includegraphics[width=0.9\textwidth]{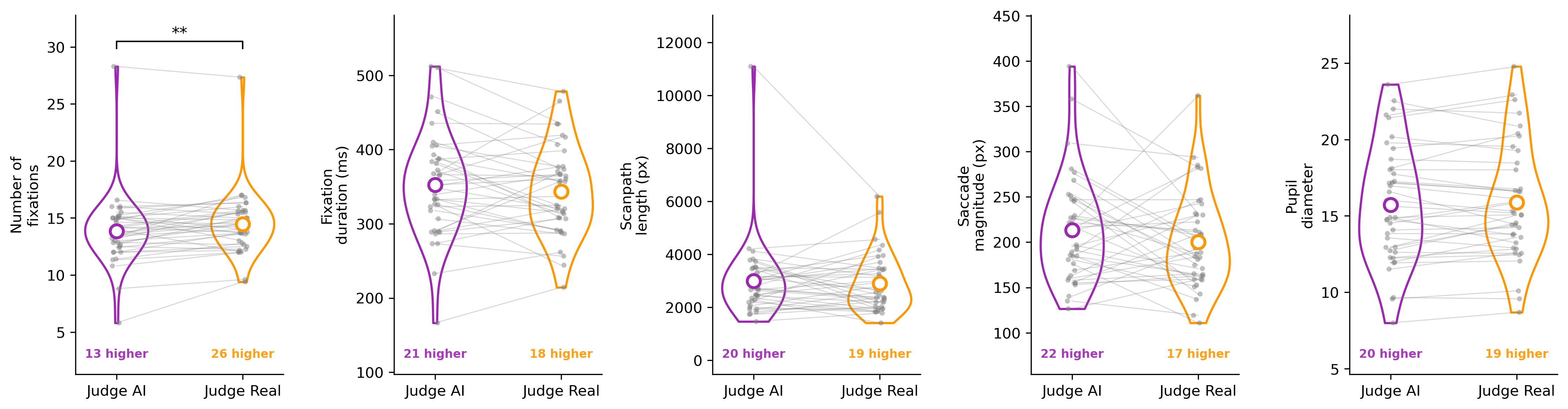}
    \caption{When watching AI-generated videos in AI detection tasks, participants who judged the videos as real significantly acquired more fixations (**).}
    \label{analysis-c}
\end{subfigure}
\caption{
Analysis of five eye-tracking metrics across experimental conditions.
}
\label{analysis}
\end{figure*}

\begin{itemize}
    \item[(1)] \textbf{Introduction (5 min).} At the start of the experiment, an instructor introduced the study to each participant. 
    \item[(2)] \textbf{PART 1: Video understanding (8 min)} 
    In the first part, participants were instructed to perform video understanding tasks (``\textit{You will watch 20 videos one after the other. Each video is 5 seconds long. After each one, you’ll be asked to describe what happened in the video in a single sentence.}'') -- they were asked to watch the videos naturally, as they would any online content, without specific goals.
    They were informed that they will be asked to describe each video in one sentence directly after watching. They complete 20 task trials (half real and half AI videos) in sequence. A black image with a fixation text in the middle appears between videos. The participant fixates on the text for three seconds before the video to standardise the gaze position.
    \item[(3)] \textbf{Break (5 min)} Participants were given a short break to recover from the first task. The instructor will let them do the next task once they feel ready to continue.
    \item[(4)] \textbf{PART 2: AI video detection (8 min)} 
    In the second part, participants were asked to perform AI detection tasks (``\textit{You will watch 20 videos one after the other. Each video is 5 seconds long. After each one, you'll be asked to determine whether the video is AI-generated. If it is, specify which visual elements led you to that conclusion}''). They were asked to watch another set of 20 videos (also half real and half AI). This time, they were instructed to: (1) watch each video carefully, (2) indicate whether they believe it is real or AI-generated, and (3) if it is AI-generated, to indicate which visual elements made the video unrealistic. The participants orally answered questions, transcribed by the experiement conductor.
    \item[(5)] \textbf{Questionnaire and Debrief (5 min)}
    After the experiment, participants completed a post-experiment questionnaire that collected general demographic information, such as age and gender, as well as the level of experience participants had with video-generation AI tools, how confident they were in their ability to identify AI-generated videos, and their strategies for identifying such content. 
\end{itemize}

\section{Results}

The core research question of this study is how people watch AI-generated videos of physical scenes. 
We initially formalize three hypotheses of human gaze behavior to be investigated in this study:

\begin{itemize}
    \item \textit{H1}: Human gaze behaviors differ when people normally watch videos compared to when they are aware and attempt to identify AI-generated content. 
    \item \textit{H2}: Human gaze behaviors differ when viewing real versus AI-generated videos.
    \item \textit{H3}: Humans with different AI-detection strategies have distinct gaze behaviors.
\end{itemize}

We analyze participant eye-tracking data and their responses. A total of 21,379 fixations were recorded across 1,573 scanpaths from 40 participants. 27 scanpaths were lost during the experiment due to hardware issues. In addition, 800 judgments were collected during the AI detection tasks, of which 796 included corresponding eye-tracking data. 
We first introduce the metrics and then divide the analysis of results into three parts based on evaluation of three hypotheses.

    
   
    


\subsection{Metrics}


We considered two classes of metrics: metrics relating to participant eye tracking data, and subjective measures into AI detection.

\paragraph{Eye tracking metrics}

\begin{itemize}
    \item \textit{Number of fixations}: The number of fixations is counted in each session. A higher number of fixations indicates more information sampling~\cite{gupta2020eyes, shi2025chartist}.
    \item \textit{Fixation duration}: The duration of the fixation in seconds. A longer fixation duration means more effortful processing at single locations~\cite{gupta2020eyes, ilyas2025reading}.
    \item \textit{Saccade magnitude}:  
    Saccade magnitude is calculated as the distance between the current fixation and the previous fixation. Greater saccade magnitude reflects shifts in attention across more distant regions, which can reflect the efficiency of visual processing.~\cite{ilyas2025reading, pannasch2008visual}.
    \item \textit{Scanpath length}:  
    Scanpath length is measured by the total distance of the user’s scanpath across the video. A longer scanpath indicates a more exploratory viewing pattern~\cite{gupta2020eyes}.
    \item \textit{Pupil size}: The diameter of the eye pupil in pixels. We use the average of left and right pupil size. Bigger pupil diameter means the increase of cognitive load as a result of central autonomic nervous system activity~\cite{szulewski2017measuring, hess1965attitude}. 
\end{itemize}

\paragraph{AI detection metrics}

\begin{itemize}
    \item \textit{Accuracy}: Human judgments are compared with the ground-truth labels for human or AI videos. This is comparable to the Turing test, where the random guess will be close to 50\% accuracy as the random baseline.
\end{itemize}

\subsection{Task-dependent gaze behaviors}

\begin{figure}[t]
  \centering
  \begin{subfigure}[t]{0.9\linewidth}
    \centering
    \includegraphics[width=0.9\linewidth]{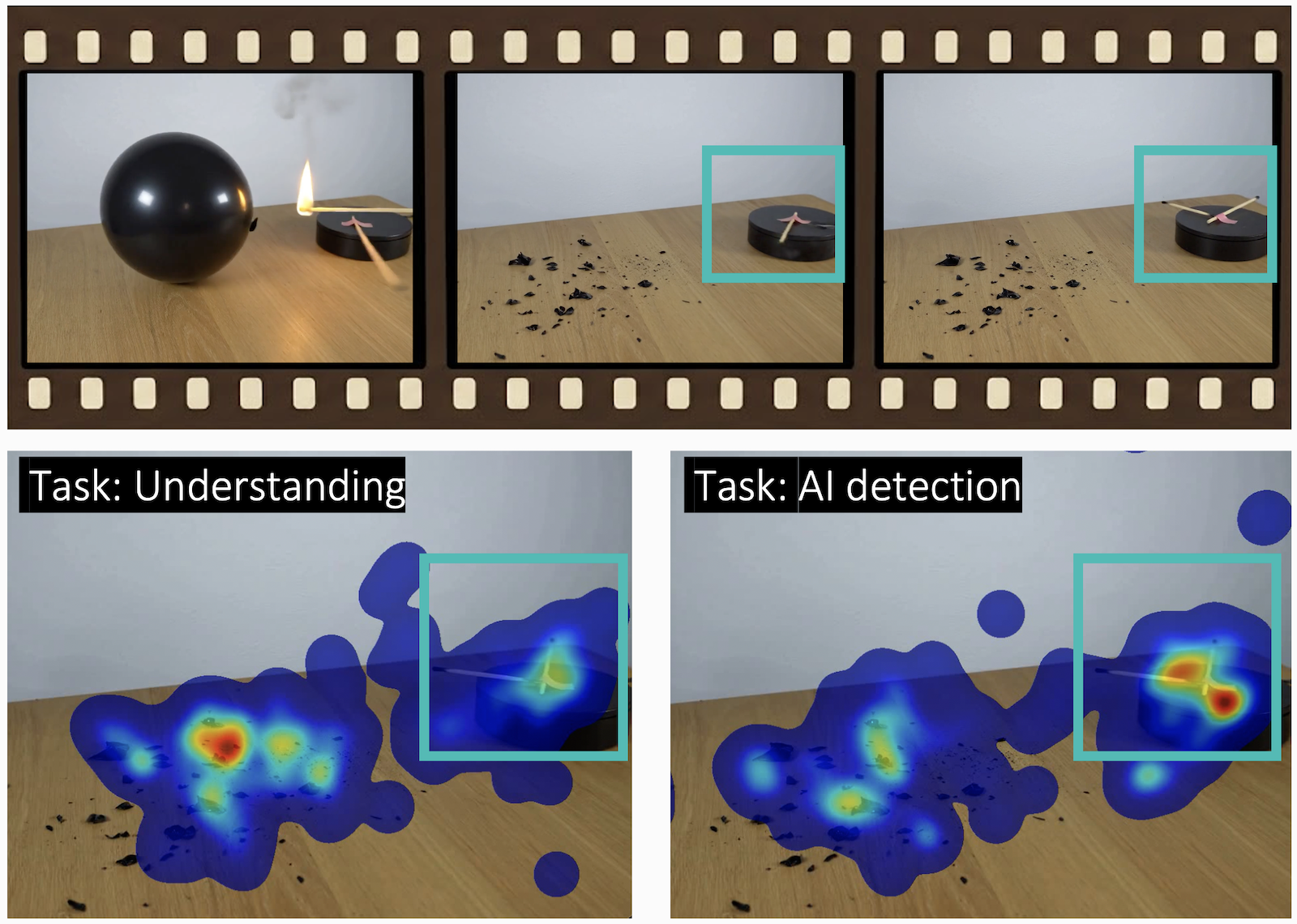}
    \caption{Humans focus more visual attention on the anomaly area (blue box) when detecting AI compared to normal viewing.}
    \label{fig:top}
  \end{subfigure}

  \vspace{0.5em}

  \begin{subfigure}[t]{0.9\linewidth}
    \centering
    \includegraphics[width=0.9\linewidth]{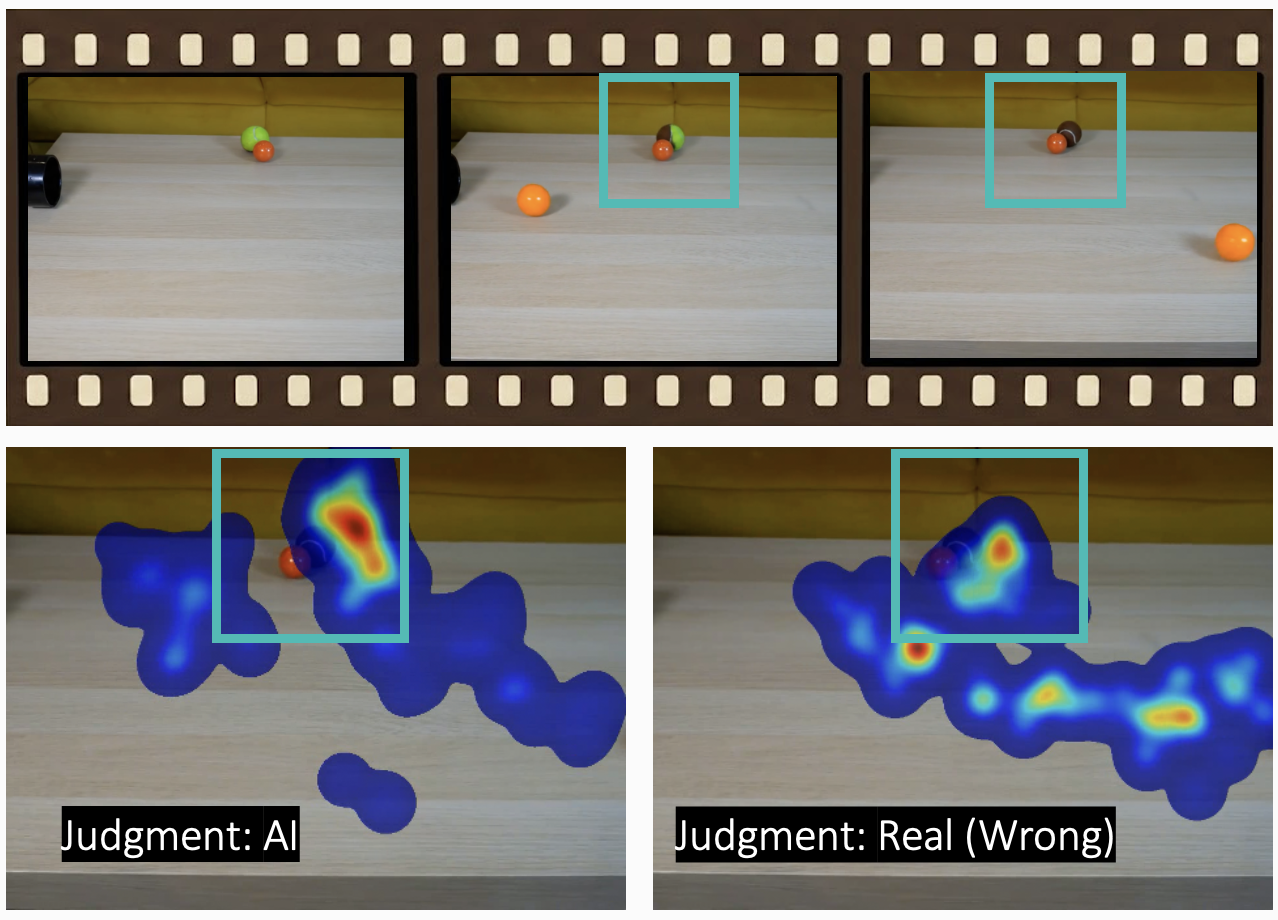}
    \caption{Humans who pay more attention to the anomaly area (blue box) have a better chance of identifying AI.}
    \label{fig:bottom}
  \end{subfigure}

  \caption{In the qualitative analysis of the aggregated human visual attentions on sample AI-generated videos, people show task-dependent and judgment-dependent gaze behaviors.}
  \label{fig:case}
\end{figure}

As illustrated in Figure~\ref{analysis} (a) and (b), participants show different gaze behavior on real-world and AI generated videos when they are asked to normally watch videos (T1) and to detect whether the videos are AI (T2).
When viewing real-world videos (Figure~\ref{analysis}-a), the mean pupil sizes were smaller during AI detection, suggesting that spotting AI required lower cognitive load than video understanding if the video is real. 
When viewing AI-generated videos (Figure~\ref{analysis}-a), participants spent more fixations but shorter fixation durations ($p<0.05$) during AI detection tasks. 
This might indicate easier processing of visual information within the AI-generated videos. 
Saccade magnitudes were longer ($p<0.05$) during AI detection tasks compared to normally watching, indicating that people shift their attention over larger areas when trying to detect AI-generated videos.
These results lend support to \textit{H1}: when people are aware of AI videos, people change their gaze behavior. They tried to rapidly sample information globally rather than processing in fewer places in the videos.

When looking at the details of human behavior during two tasks, people actively look for anomalies and pay less attention to what is happening in the video when they are aware of AI. 
For instance, as shown in Figure~\ref{fig:case}-a, it is an AI-generated video showing a balloon being poked by a firing stick and then bursting, which is the most eye-catching event in the video. At the same time, there is a rotation apparatus where the stick changes from one to two, which is an error caused by AI generation.
For people normally watching videos for understanding, they focus mostly on the key event, which is the burst of the balloon, and then look at the residues on the table. 
However, the people who are trying to detect AI have more distributed attention and focus more on the anomaly area where the error occurs.



\subsection{Judgment-dependent gaze behaviors}

\begin{figure}[!t]
    \centering

    \includegraphics[width=0.95\linewidth]{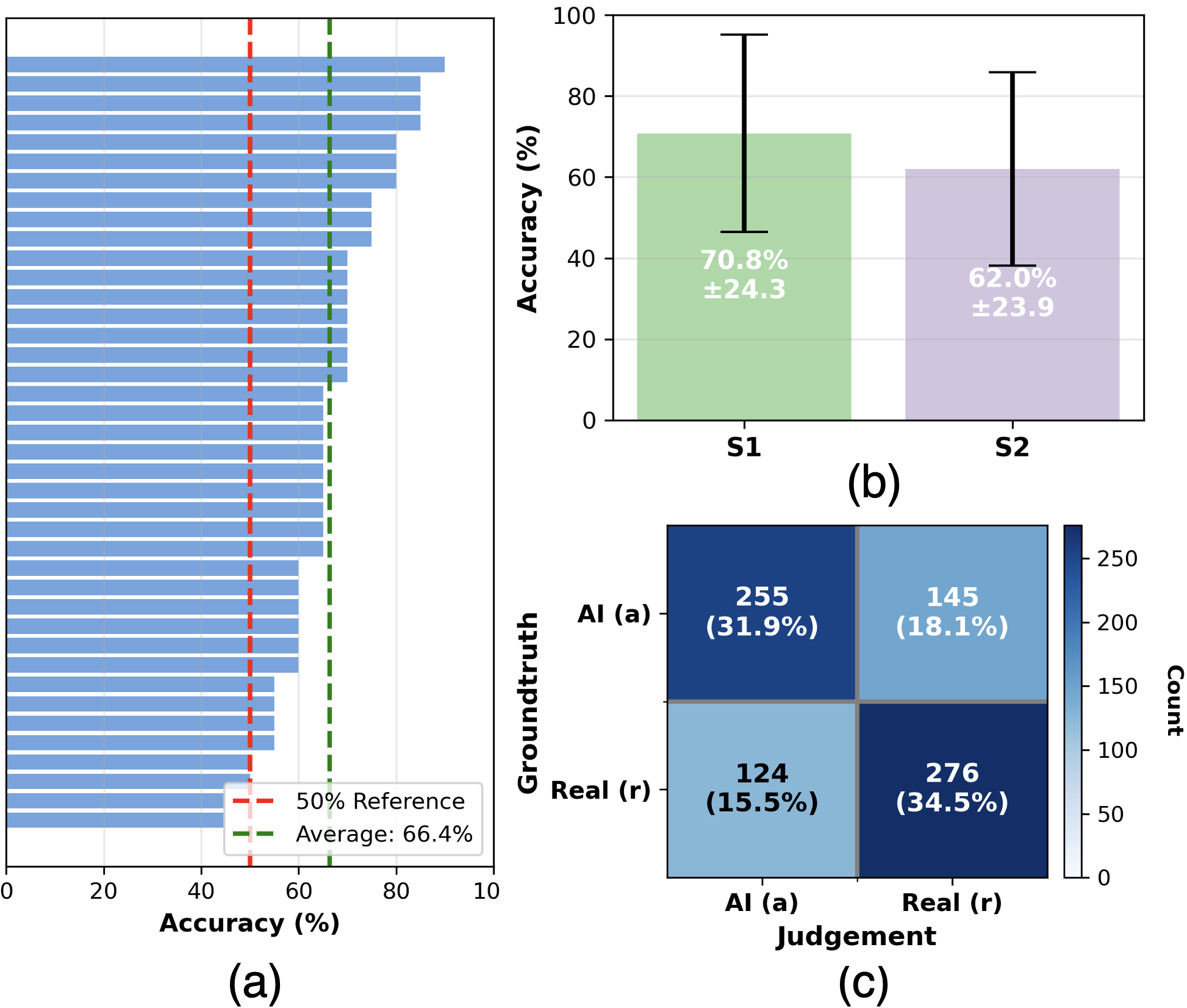}
    \caption{Accuracy of AI detection: (a) User accuracy ranking of 40 participants; (b) Accuracy comparison between S1, physics videos, and S2, professional videos; (c) Confusion matrix. We see more errors due to judging real videos as AI.} 
    \label{fig:judgments}
    \vspace{-3mm}
\end{figure}

Figure~\ref{fig:judgments} presents the judgment accuracies of all participants in detecting AI-generated content. The majority of participants (36 out of 40) performed better than the random guess baseline of 50\%. The mean accuracy across participants was 66.4\%.
A comparison between the two video sets reveals that participants were more successful at detecting AI-generated videos in S1, physics videos, (M = 70.8\%, SD = 24.3\%), compared to S2, professional videos (M = 62.0\%, SD = 23.9\%).
The confusion matrix indicates that a higher proportion of AI-generated videos were misclassified as real (18.1\%) compared to real videos misclassified as AI-generated (15.5\%).

Comparing the gaze patterns when participants viewed AI-generated and real-world videos revealed almost no significant differences between gaze patterns in either task (Figure~\ref{analysis}-c).  
The statistically significant difference occurred during the AI detection task, where participants exhibited larger pupil sizes while viewing AI-generated videos compared to real ones ($p < 0.05$). 
This finding might suggest that participants put more cognitive effort into evaluating AI-generated videos. 
However, in the present study, we did not control the luminance of the AI-generated videos, which may also cause pupils to dilate.






Although participants' gaze behavior did not depend on the true nature of the video (AI or real), differences did emerge based on participants' \emph{judgments} regarding the video's authenticity (Figure~\ref{analysis}-d).
Participants had significantly more fixations ($p<0.01$) but shorter fixation durations (not significant, $p=0.28$) when they thought that an AI-generated video was real. 
This suggests that participants scanned the video trying to spot anomalies, a process that continues until the viewer (mistakenly) judges the video as real. 
This lends partial support to \textit{H2}. 
People exhibit different gaze behaviors when watching AI-generated versus real videos, with particular difference primarily when they perceive a video as real or AI-generated.

People who made incorrect judgments about AI content show different attention than those who made correct judgments. 
As illustrated in Figure~\ref{fig:case}-b, the AI-generated video contains three balls. One orange ball rapidly rolls from left to right, while a tennis ball subtly changes color from green to brown, representing the anomaly.
A comparison of aggregated visual attention between participants who identified the video as AI-generated and those who believed it was not reveals that people who failed to detect the AI-generated content focused more on areas outside the anomaly. 
This pattern explains their failure to identify the anomaly and their significantly higher number of fixations.





\subsection{Strategy-based gaze behaviors}

\begin{figure}[!t]
\centering
\includegraphics[width=\linewidth]{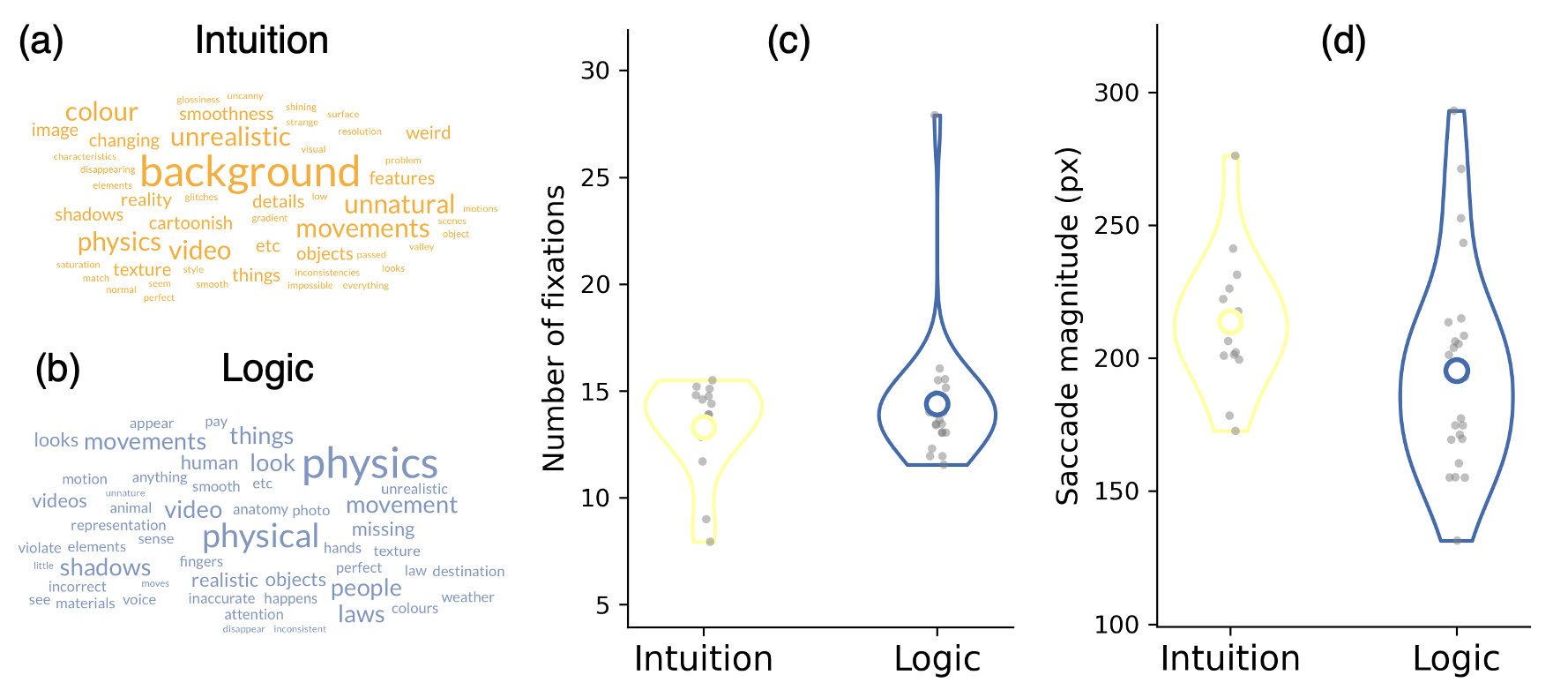}
\caption{
Participants are manually grouped into two categories based on their self-reported startegies: (a) \textit{intuition}, and (b) \textit{logic}.
Gaze behavior has different patterns in (c) number of fixations and (d) saccade magnitude between these two general strategies in spotting AI. Participants with a \textit{logic} strategy have a more targeted attention distribution.
} 
\label{fig:strategies}
\end{figure}

During the experiment, we asked participants to summarize their strategies for distinguishing between real and AI-generated videos. Three of the authors reviewed the comments from 40 participants and reached a consensus that the strategies can be grouped into two main categories: \textit{intuition} and \textit{logic} (see the tag clouds of frequently mentioned words in Figure~\ref{fig:strategies}-a and b). 
\textit{Logic} refers to participants trying to spot anomalies and identify places where the video does not align with their model of the physical world, relating to comments such as ``\textit{Defy nature law of physics}'' and ``\textit{The stability of the element}''. 
On the other hand, participants following an \textit{Intuition} strategy, describe a general feeling that the video is AI-generated, with comments such as ``\textit{Artificial looking textures}'' and ``\textit{Everything looks too perfect}''.

We label the participants' strategies and further analyse thier gaze behavior based on the two strategies during the AI detection task (Logic: 25 participants; Intuition: 15 participants).
We find there are significant differences in the number of fixations and saccade magnitudes, as illustrated in Figure~\ref{fig:strategies}-c and d. Participants who employed the logical strategy sampled more positions on the videos, but had shorter saccade magnitudes, indicating that they engaged in a more logical exploration guided by a targeted attention distribution.
These results lend support to \textit{H3}. 
Human gaze behaviors during video watching vary in their AI-detection strategies. Future research can explore more detailed classifications of human strategies.

\section{Discussion} 

In this study, we investigated differences in eye movement behavior when watching AI-generated and real videos. We find that the authenticity of the video, i.e., whether it is real or AI-generated, does not have a significant effect on gaze behavior. However, the act of attempting to detect whether or not a video is fake (compared to normal viewing) does change eye gaze behavior. Importantly, participants' judgment of whether or not the video is real (irrespective of the truth) altered their gaze behavior.   

Interestingly, some of our findings diverge from previous studies. An investigation of DeepFake face-swapped videos~\cite{gupta2020eyes} identified significant differences in viewing patterns between real and AI-generated face-swapped videos, with a greater number of fixations and longer scanpaths observed for real videos. This difference could be because humans are particularly attuned to visual details in human faces. However, the findings from our study could suggest that those differences were due to participants' judgments rather than whether the stimulus video was indeed AI-generated or not. Given the increasingly high fidelity of modern generative AI models, our findings suggest that perception matters more than reality when the ground truth is visually ambiguous. When participants assess whether a video is AI-generated, their expectations influence their viewing behavior, prompting them to search for anomalies to confirm their hypothesis. 

These findings have significant overarching implications -- they suggest that our behavior changes not only when we engage with an AI-generated video, but as soon as we are aware that videos \textit{can} be fake. With the increasing prevalence and growing awareness of AI-generated content, this is likely to change how people interact with the video medium altogether. While one can argue this is true for other generated content, videos have been, for years, considered one of the most reliable forms of evidence, underpinning critical societal processes from fair trials to democratic discourse. While it is well-known that generative AI endangers the reliability of evidence, our results highlight a secondary, perhaps more insidious risk: the mere possibility of AI fabrication already changes how we witness and interact with the world.

\bibliographystyle{apacite}

\setlength{\bibleftmargin}{.125in}
\setlength{\bibindent}{-\bibleftmargin}

\bibliography{CogSci}

@misc{google2025veo3,
  author       = {Google DeepMind},
  title        = {Veo 3: Video generation model},
  year         = {2025},
  howpublished = {\url{https://deepmind.google/veo/}},
  note         = {Accessed: October 2025}
}

@misc{adobestock,
  author       = {Adobe},
  title        = {Adobe Stock: photos, images, graphics, vectors \& videos},
  year         = {2025},
  howpublished = {\url{https://stock.adobe.com}},
  note         = {Accessed: October 2025}
}

@misc{isthisAI,
  author       = {Reddit},
  title        = {r/isthisAI},
  year         = {2025},
  howpublished = {\url{https://www.reddit.com/r/isthisAI/}},
  note         = {Accessed: October 2025}
}

@article{motamed2025generative,
  title={Do generative video models understand physical principles?},
  author={Motamed, Saman and Culp, Laura and Swersky, Kevin and Jaini, Priyank and Geirhos, Robert},
  journal={arXiv preprint arXiv:2501.09038},
  year={2025}
}

@inproceedings{teng2025eye,
  title={Eye movement behavior during mind wandering in older adults},
  author={Teng, Xiaoru and Wong, Gloria and Chan, Antoni B and Hsiao, Janet},
  booktitle={Proceedings of the Annual Meeting of the Cognitive Science Society},
  volume={47},
  year={2025}
}

@inproceedings{teng2024eye,
  title={Eye Movement Behavior during Mind Wandering across Different Tasks in Interactive Online Learning},
  author={Teng, Xiaoru and Lan, Hui and Wong, Gloria and Chan, Antoni B and Hsiao, Janet},
  booktitle={Proceedings of the Annual Meeting of the Cognitive Science Society},
  volume={46},
  year={2024}
}

@inproceedings{hsiao2020role,
  title={The role of eye movement consistency in learning to recognise faces: Computational and experimental examinations},
  author={Hsiao, Janet H and An, Jeehye and Chan, Antoni B},
  booktitle={Proceedings of the Annual Meeting of the Cognitive Science Society},
  volume={42},
  year={2020}
}

@article{szulewski2017measuring,
  title={Measuring physician cognitive load: validity evidence for a physiologic and a psychometric tool},
  author={Szulewski, Adam and Gegenfurtner, Andreas and Howes, Daniel W and Sivilotti, Marco LA and Van Merri{\"e}nboer, Jeroen JG},
  journal={Advances in Health Sciences Education},
  volume={22},
  number={4},
  pages={951--968},
  year={2017},
  publisher={Springer}
}

@article{hess1965attitude,
  title={Attitude and pupil size},
  author={Hess, Eckhard H},
  journal={Scientific american},
  volume={212},
  number={4},
  pages={46--55},
  year={1965},
  publisher={JSTOR}
}

@inproceedings{huang2025analysis,
  title={Analysis of human perception in distinguishing real and ai-generated faces: An eye-tracking based study},
  author={Huang, Jin and Gopalakrishnan, Subhadra and Mittal, Trisha and Zuena, Jake and Pytlarz, Jaclyn},
  booktitle={2025 IEEE 19th International Conference on Automatic Face and Gesture Recognition (FG)},
  pages={1--10},
  year={2025},
  organization={IEEE}
}

@inproceedings{ilyas2025reading,
  title={Reading the Readers Mind through Eye Tracking: Can AI Generated Texts Match Human Authors?},
  author={Ilyas, Chaudhary Muhammad Aqdus and Noor, Sifat-E and Tashk, Ashkan and Cooreman, Bart and Beier, Sofie and B{\ae}kgaard, Per},
  booktitle={Proceedings of the 2025 Symposium on Eye Tracking Research and Applications},
  pages={1--7},
  year={2025}
}

@article{cartella2024unveiling,
  title={Unveiling the truth: Exploring human gaze patterns in fake images},
  author={Cartella, Giuseppe and Cuculo, Vittorio and Cornia, Marcella and Cucchiara, Rita},
  journal={IEEE Signal Processing Letters},
  volume={31},
  pages={820--824},
  year={2024},
  publisher={IEEE}
}

@inproceedings{wohler2021towards,
  title={Towards understanding perceptual differences between genuine and face-swapped videos},
  author={W{\"o}hler, Leslie and Zembaty, Martin and Castillo, Susana and Magnor, Marcus},
  booktitle={Proceedings of the 2021 CHI conference on human factors in computing systems},
  pages={1--13},
  year={2021}
}

@inproceedings{gupta2020eyes,
  title={The eyes know it: Fakeet-an eye-tracking database to understand deepfake perception},
  author={Gupta, Parul and Chugh, Komal and Dhall, Abhinav and Subramanian, Ramanathan},
  booktitle={Proceedings of the 2020 international conference on multimodal interaction},
  pages={519--527},
  year={2020}
}

@article{pannasch2008visual,
  title={Visual fixation durations and saccade amplitudes: Shifting relationship in a variety of conditions},
  author={Pannasch, Sebastian and Helmert, Jens R and Roth, Katharina and Herbold, Ann-Katrin and Walter, Henrik},
  journal={Journal of Eye Movement Research},
  volume={2},
  number={2},
  year={2008}
}

@inproceedings{shi2025chartist,
  title={Chartist: Task-driven Eye Movement Control for Chart Reading},
  author={Shi, Danqing and Wang, Yao and Bai, Yunpeng and Bulling, Andreas and Oulasvirta, Antti},
  booktitle={Proceedings of the 2025 CHI Conference on Human Factors in Computing Systems},
  pages={1--14},
  year={2025}
}

@article{renaultgrok,
  title={@ Grok Is This True? LLM-Powered Fact-Checking on Social Media},
  year         = {2025},
  author={Renault, Thomas and Mosleh, Mohsen and Rand, David},
  publisher={OSF}
}

@misc{sora,
  author       = {OpenAI},
  title        = {Sora is here},
  year         = {2025},
  howpublished = {\url{https://openai.com/index/sora-is-here/}},
  note         = {Accessed: October 2025}
}

@inproceedings{shi2024crtypist,
  title={CRTypist: Simulating touchscreen typing behavior via computational rationality},
  author={Shi, Danqing and Zhu, Yujun and Jokinen, Jussi PP and Acharya, Aditya and Putkonen, Aini and Zhai, Shumin and Oulasvirta, Antti},
  booktitle={Proceedings of the 2024 CHI Conference on Human Factors in Computing Systems},
  pages={1--17},
  year={2024}
}

@inproceedings{tanaka2019utilizing,
  title={Utilizing eye-tracking to explain variation in response to inconsistent message onbelief change in false rumor},
  author={Tanaka, Yuko and Inuzuka, Miwa and Hirayama, Rumi},
  booktitle={Proceedings of the Annual Meeting of the Cognitive Science Society},
  volume={41},
  year={2019}
}

@article{westerlund2019emergence,
  title={The emergence of deepfake technology: A review},
  author={Westerlund, Mika},
  journal={Technology innovation management review},
  volume={9},
  number={11},
  year={2019}
}

\end{document}